# Extrinsic Photodiodes for Integrated Mid-Infrared Silicon Photonics

Richard R. Grote,[1,†] Brian Souhan,[1] Noam Ophir,[1] Jeffrey B. Driscoll,[1] Keren Bergman,[1] Hassaram Bakhru,[2] William M. J. Green,[3] and Richard M. Osgood, Jr.[1]

[1]Department of Electrical Engineering, Columbia University, 500 W. 120th Street, New York, NY 10027, USA
[2]College of Nanoscale Science and Engineering, State University of New York at Albany, 257 Fuller Road, Albany, NY 12203, USA
[3]IBM T. J. Watson Research Center, 1101 Kitchawan Rd., Yorktown Heights, NY 10598, USA
[†]Current affiliation: Department of Electrical and Systems Engineering, University of Pennsylvania, 200 S. 33rd Street, Philadelphia, PA 19104, USA

**Silicon photonics has recently been proposed for a diverse set of applications at mid-infrared wavelengths[1,2], the implementation of which require on-chip photodetectors. In planar geometries, dopant-based extrinsic photoconductors have long been used for mid-infrared detection with Si and Ge acting as host materials[3]. Leveraging the dopant-induced sub-bandgap trap-states used in bulk photoconductors for waveguide integrated mid-infrared detectors offers simple processing, integration, and operation throughout the mid-infrared by appropriate choice of dopant (Fig. 1a). In particular, Si doped with Zn forms two trap levels $\approx$ 0.3 eV and $\approx$ 0.58 eV above the valence band[4-6], and has been utilized extensively for cryogenically cooled bulk extrinsic photoconductors[3,7]. In this letter, we present room temperature operation of $Zn^+$ implanted Si waveguide photodiodes (Figs. 1b,c) from 2.2 μm to 2.4 μm, with measured responsivities of up to 87 ± 29 mA/W and low dark currents of < 10μA.**

While a number of Si waveguide (SiWG) integrated optoelectronic devices have been demonstrated in this wavelength range[8,9], efficient photodetection remains an important and challenging task. Thus, spectral translation of mid-infrared signals to the telecom regime via four-wave mixing in SiWGs has been proposed for on-chip detection[10], which makes use of the sensitive integrated photodiodes (PDs) in the telecommunications wavelength range[11]. However, this method requires a high-powered pump laser and long, on-chip-waveguide lengths to achieve efficient wavelength conversion. In addition, heterogeneous integration of both narrow-bandgap semiconductors[9,12-14] and graphene[15,16] with SiWGs has been demonstrated for on-chip mid-infrared detection. Though viable PDs have been demonstrated, heterogeneous integration schemes present an inherent difficulty by imposing constraints on material quality and process integration.

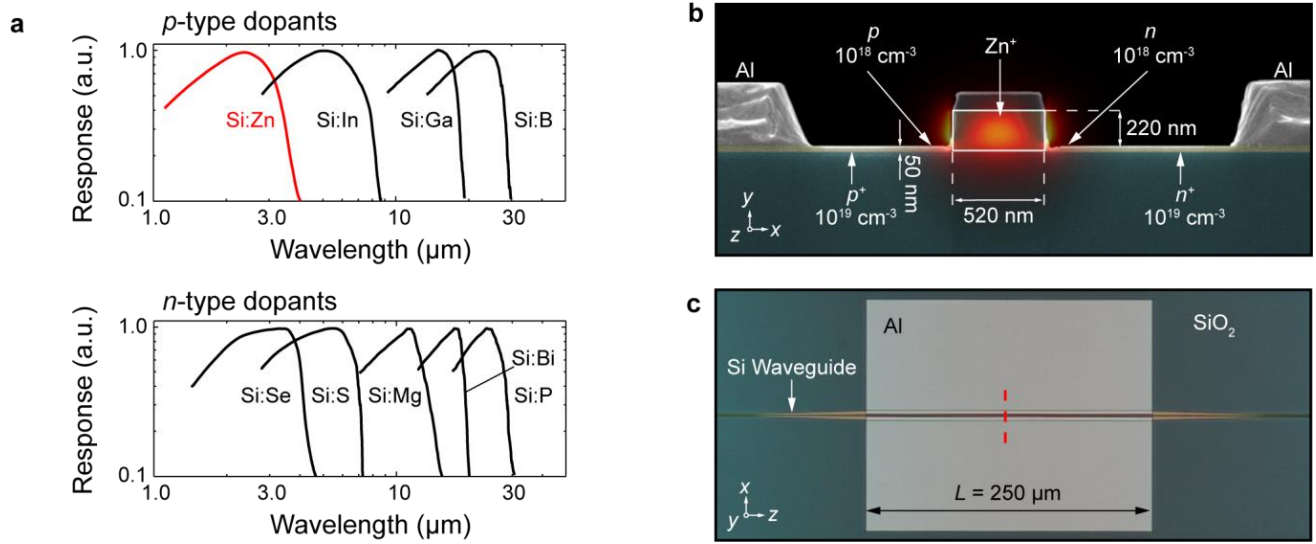

**Fig. 1 | a** Normalized response of bulk silicon photoconductors for different dopants, re-plotted from ref. 3. Dopants are designated as either *p*-type or *n*-type depending upon whether the resultant trap state is closer to the valence band or conduction band, respectively. **b** False color SEM cross-section of the *p*-Si:Zn-*n* PD with finite-element-method calculated quasi-TE mode intensity at a wavelength of 2.3 μm superimposed. The waveguide has a 90 nm-thick $SiO_2$ hardmask and 3 μm-thick buried-oxide-layer substrate. **c** Top-view optical microscope image of the PD. The red dashed line indicates the position of the SEM cross-section shown in **b**.

Extrinsic detectors, which utilize absorption transitions from dopant-induced trap states within the bandgap of a host material, present a simple solution for high-performance integrated mid-infrared PDs and alleviate the need for heterogeneous integration. These PDs can potentially be integrated into a standard CMOS process flow by adding an ion implantation and annealing step after activation of the source and drain implants, and prior to the deposition of back-end dielectrics and interconnect metallization. Alternatively, this additional fabrication step can be performed as a post-process, as is done here (see Methods). The Si dopants used in bulk photoconductors can potentially be integrated for detection from a range of wavelengths from 1.5 μm to greater than 25 μm (Fig. 1**a**), while doped Ge bulk photoconductors have been demonstrated at wavelengths greater than 100 μm[3].

For the 2.2 μm to 2.4 μm wavelength range Zn, Se, and Au dopants have been shown to produce a suitable defect level in Si[5], and planar photoconductive detectors based upon these dopants have been demonstrated[3,7]. While these detectors required liquid nitrogen cooling, it has very recently been shown that silicon hyperdoped with Au can generate photocurrent up to a wavelength of 2.2 µm at room temperature with a Schottky contact configuration[17]. Since cooling is not required to eliminate dark current in diode-based detectors, room temperature operation is achieved with the PDs explored here as well. Although transition metals such as Au and Zn are generally not CMOS compatible, as they can diffuse into Si and adversely affect carrier mobility, the other dopants in Fig. 1**a** do not have these restrictions. For the wavelength range explored here, Si:Se could be substituted for CMOS compatibility, and has been demonstrated for a SiWG PD operating at 1.55 µm[18].

The PDs demonstrated here are based on a *p-i-n* diode structure fabricated in a 250 µm long Si rib/ridge waveguide with 520 nm × 220 nm channel section and a 50 nm doped silicon ridge used to form ohmic contacts to Al (see Methods), as shown in Figs. 1**b,c**. The intrinsic region of the *p-i-n* diode corresponds to the channel section of the SiWG, and is implanted with $Zn^+$ to form the final *p*-Si:Zn-*n* PD structure. Two $Zn^+$ implantation doses are investigated, $10^{12}$ cm$^{-2}$ and $10^{13}$ cm$^{-2}$, corresponding to estimated average Zn concentrations inside the channel section of the waveguide of $N_{Zn}$ = 4.5×10$^{16}$ cm$^{-3}$ and $N_{Zn}$ = 4.5×10$^{17}$ cm$^{-3}$, respectively. The Zn concentrations are estimated by averaging the stopping range of $Zn^+$ inside of the Si device layer, which is calculated by SRIM software[19] for an acceleration voltage of 260 keV. The acceleration voltage is chosen for maximum overlap between the generated trap states and the quasi-TE waveguide mode (see SI). Subsequent to implantation, the PDs were annealed in atmosphere for a series of increasing temperatures, reaching a maximum of 350°C, and the responsivity was found to increase with annealing temperature.

The detection mechanism of our *p*-Si:Zn-*n* PDs is due to substitutional Zn atoms in the Si lattice, which act as a double-acceptors and result in the two defect levels (Fig. 2**a**) with energy levels of $E_{d1} \approx E_v + 0.3$ eV and $E_{d2} \approx E_v + 0.58$ eV[5,6]. While the position of the Fermi level in the Si:Zn region is not known, the excess carrier concentration is below that of the *p* and *n* regions, ensuring that the Si:Zn region will be fully depleted with the application of a reverse bias voltage. Photon-induced transitions occurs between $E_{d2}$ and the conduction band, which corresponds to a transition energy of $E_g - E_{d2} = 1.12\text{eV} - 0.58\text{eV} = 0.54\text{eV}$ with a peak absorption wavelength of ≈ 2.3 µm. Photocurrent generation due to this transition is shown to be a single-photon process by the linearity measurements in Fig. 2**b**. The presence of $E_{d1}$ does not contribute to photocurrent generation in the wavelength range of interest; however, its presence should impact the thermally assisted re-population rate of $E_{d2}$ and thus the internal quantum efficiency, $\eta_i$, of the PD.

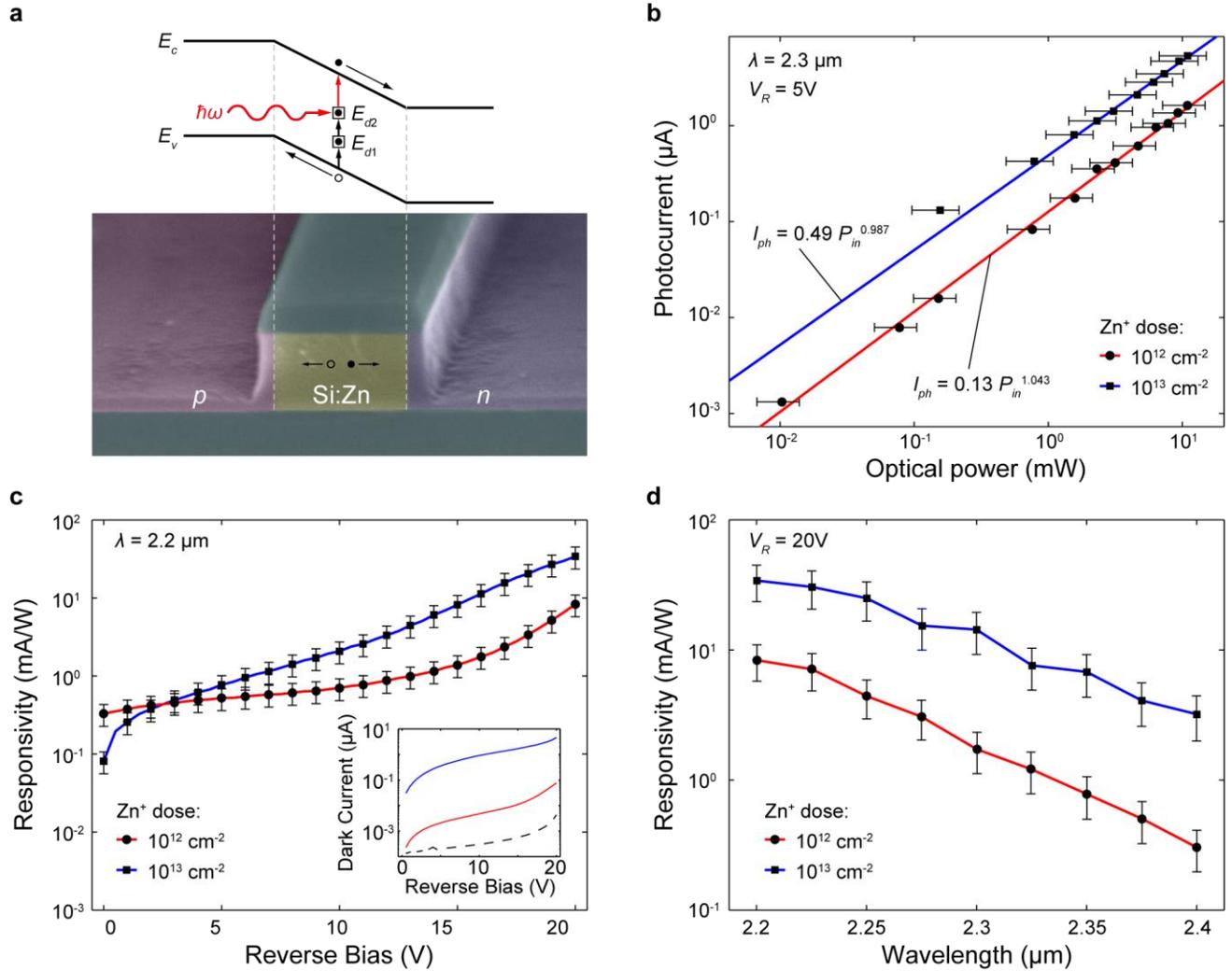

**Fig. 2** | **a** Band diagram of the *p*-Si:Zn-*n*, photodiode with defect levels[6] $E_{d1} \approx E_v + 0.3$ eV and $E_{d2} \approx E_v + 0.58$ eV in the Si:Zn region. The doping profile is illustrated by the false-color SEM shown below. **b** Photocurrent $I_{ph}$ versus input optical power $P_{in}$ along with linear fits. The linear correspondence indicates a single-photon excitation process. The higher implantation dose is shown for a 3 mm length PD. **c** Responsivity as a function of reverse bias voltage. Inset: dark current as a function of reverse bias voltage for an un-implanted diode (black dashed line) and PDs with Zn$^+$ doses of $10^{12}$ cm$^{-2}$ (red solid line) and $10^{13}$ cm$^{-2}$ (blue solid line). **d** Responsivity as a function of wavelength. The decrease in responsivity with increasing wavelength is due to parasitic absorption. Error bars have been determined by variances in system losses, facet loss, and fluctuations in transmitted power (see Methods).

The responsivity, defined as $R = I_{ph}/P_{in}$, where $I_{ph}$ is the photocurrent and $P_{in}$ is the on-chip optical power entering the PD, is measured as a function of reverse bias voltage with 2.2 μm-wavelength excitation (Fig. 2**c**). While the dark current (Fig. 2**c**, inset) increases with dopant concentration, it remains below 10 µA for all devices even at the maximum reverse bias voltage of 20V under which measurements were taken. The kink in the responsivity curve of the $10^{12}$ cm$^{-2}$ implantation dose PDs between 12V and 15V is indicative of avalanche multiplication, which has previously been observed in similar SiWG *p-i-n* contact configurations[20]. The responsivity is largest at shorter wavelengths (Fig. 2**d**), reaching a maximum of 8.3 ± 2.6 mA/W and 34.1 ± 10.6 mA/W for doses of $10^{12}$ cm$^{-2}$ and $10^{13}$ cm$^{-2}$, respectively, at a wavelength of 2.2 µm and a reverse bias of 20V. For 3 mm-length PDs maximum responsivities of 87 ± 29 mA/W were measured under similar conditions, and the dark current remained below 10µA at 20V reverse bias (see SI). The decrease in responsivity with increasing wavelength is due to parasitic absorption from the contacts (free-carrier absorption in *p, n* doped regions as well as the Al regions), which increases from > 30% at 2.2 µm to > 89% at 2.4 µm for a dose of $10^{12}$ cm$^{-2}$, and from > 10% at 2.2 µm to > 85% at 2.4 µm for a dose of $10^{13}$ cm$^{-2}$ (see SI). The parasitic absorption has been determined by transmission measurements through an un-implanted diode with the same geometry, *p, n* doping, and contact metallization (see Methods). By moving the contacts and *p, n* doped regions further from the waveguide, the responsivity at longer wavelengths can be substantially improved.

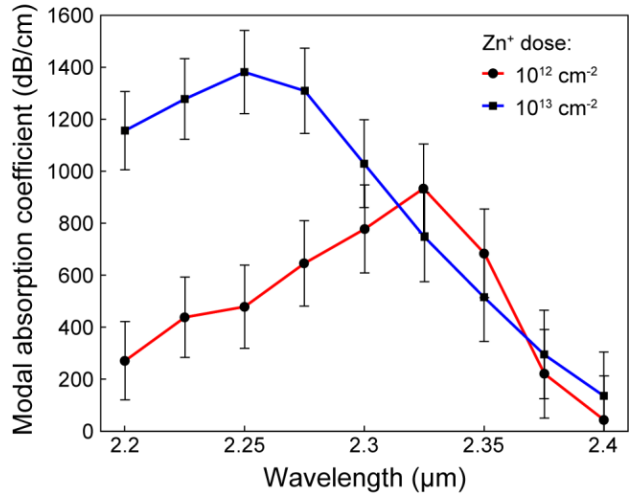

**Fig. 3** | Modal absorption coefficients due to Zn$^+$ implantation. Parasitic contact absorption has been normalized out, and error bars have been determined by variances in system losses, facet loss, and fluctuations in transmitted power (see Methods).

The modal absorption due to Zn$^+$ implantation is determined by transmission measurements with parasitic contact absorption being taken into account (see SI), and is shown in Fig. 3 for both doses. For the PD with a Zn$^+$ dose of $10^{12}$ cm$^{-2}$, the modal absorption coefficient peaks at a wavelength of 2.325 µm corresponding to a transition energy of 0.533 eV. Similarly, the PD with a $Zn^+$ dose of $10^{13}$ $cm^{-2}$ exhibits a peak in the modal absorption coefficient at a wavelength of 2.250 µm, corresponding to a transition energy of 0.551 eV. These peak absorption wavelengths correspond to the transition from $E_{d2}$ to the conduction band shown in Fig. 2**a**.

Even with parasitic absorption taken into account, the modal absorption coefficient along with the measured responsivities indicate that $\eta_i$ of the PDs is below 5%, which is significantly lower than the reported values of $\eta_i = 20\%$ at a wavelength of 2.4 µm for Si:Zn bulk photoconductive detectors fabricated by diffusion doping[7]. Though diffusion doping of Zn in Si results in electrical activation by forming substitutional defects on Si lattice sites[7], activation of the $Zn^+$-implanted Si used here is not well characterized. For $Zn^+$ implantation doses above the amorphization threshold, Zn out-diffusion and Si recrystallization occur concurrently between 300°C and 500°C annealing[21, 22], and a significant fraction of Zn atoms remain on interstitial lattice sites throughout the process[22]. Although the $Zn^+$ doses used are below the amorphization threshold dose[23] of $3\times10^{14}$ $cm^{-2}$ (see SI), Si lattice defects caused by implantation are expected to remain for annealing temperatures below 550°C[22]. The post-implantation annealing temperature used here was limited to 350°C to avoid reflow of the Al contacts, suggesting that there are two factors limiting $\eta_i$ of the $p$-Si:Zn-$n$ PDs: (1) lattice defects that remain in the Si lattice after annealing, and (2) the fraction of implanted Zn atoms that have not been activated.

Though certain lattice defects can contribute to photocurrent generation, such as the Si di-vacancy defect which has been used extensively for SiWG PDs[20,24-27], these particular lattice defects anneal out at temperatures above 300°C[24]. Furthermore, the responsivity of Si di-vacancy PDs decreases rapidly with increasing wavelength[27], as the position of the trap state has been identified to lie at $E_{divacancy} = E_c - 0.4$ $eV^{28}$. Thus, the absorption coefficients measured in Fig. 3 and the post-implantation annealing temperature used indicate contributions to photocurrent by Si di-vacancies is not significant. However, the remaining lattice defects are potentially contributing to optical scattering and increased carrier recombination.

While the number of activated Zn dopants in the PDs is not known, the responsivity was found to increase with increasing annealing temperatures (see Methods), suggesting that activation can be improved further with higher-temperature annealing. Additional implantation parameters can also be optimized, as improved activation and reduced out-diffusion have been achieved by performing Zn$^+$ implantation at 350°C[22]. Alternatively, other doping methods can be used to produce higher activation such as pulsed laser melting to supersaturate the Si with Zn[29]. Beyond processing improvements for increased Zn activation, $\eta_i$ can potentially be improved by optimizing the thermal repopulation rates of the trap-states to increase the probability of a photon absorption event. The repopulation rates are determined by the Fermi level and can be engineered via compensation doping with conventional *p* and *n* shallow donors[28,30].

Our results illustrate the potential for silicon photonic extrinsic photodiodes operating at mid-infrared wavelengths. The use of dopants for trap-state engineering in diode configurations allows for design of on-chip PDs with the device parameters such as dark current, internal quantum efficiency, and peak detection wavelength being controlled through adjustments to the trap-state levels and populations. Beyond photodetection, trap-state transitions can emit photons under the proper conditions[31], creating the prospect for on-chip waveguide-integrated sources and detectors operating throughout the mid-infrared wavelength band.

**Methods:**

**Device fabrication.** Un-implanted *p-i-n* SiWG diodes were fabricated on the CMOS line at MIT Lincoln Laboratory as described in ref. 26 using SOI with a 3 μm BOX and with the device dimensions given in Fig. 1**a**. Subsequently, contact photolithography was used to define a Shipley S1811 implantation mask at Brookhaven National Laboratory, and Zn$^+$ ion implantation was performed in the Ion Beam Laboratory, State University of New York at Albany. An acceleration voltage of 260 keV was used, corresponding to a stopping range of 118 nm into the waveguide as calculated by SRIM software[19], resulting in the maximum overlap between the induced defect concentration and the fundamental quasi-TE mode of the waveguide (see SI). The ion beam was launched 7° from normal incidence to the sample surface, and implantation current densities of 1 nA/cm² and 3-7 nA/cm² were used for the $10^{12}$ cm$^{-2}$ and $10^{13}$ cm$^{-2}$ implantation doses, respectively. Post implantation, the devices were annealed in atmosphere at 250°C for 10 minutes. Subsequently, a series of anneals were performed with increasing temperatures in 50°C steps for 10 minutes per step. Photocurrent measurements were taken between each step, and the device performance was found to improve with each annealing cycle. The annealing temperature was limited to 350°C, since the 100 nm thick Al contacts were found to reflow above this temperature.

**Experimental setup.** Optical power from an external-cavity $Cr^{2+}$:ZnSe tunable laser (IPG Photonics) is coupled into the SiWG PD using a lensed-tapered fiber (Oz Optics) to a 5μm × 220 nm silicon fan-out taper. Bias voltages are applied and currents are measured using a Keithley 2400 Source/Meter, and the transmitted power through the PD was monitored on a Yokogawa AQ6375 optical spectrum analyzer with the same coupling scheme as the input. The on-chip power entering the PD is determined by measuring the laser output power at each wavelength, and subtracting the loss from propagation through the input lensed-tapered fiber, as well as facet loss from coupling between the lensed-tapered fiber and the silicon fan-out taper. A schematic of the experimental setup is shown in the supporting information.

**Determination of error bars.** Error bars are determined by uncertainties in losses from the measurement setup (see SI), which include lensed-tapered fiber (LTF) losses and single-mode fiber transmission/bending losses. Additionally, uncertainty in the waveguide facet loss from fan-out tapers couplers have also been taken into account and as well as temporal fluctuations in the laser output as tracked by the transmitted power. Further details are provided in the supporting information.

**Extraction of the modal absorption coefficient**. Transmission measurements were performed $Zn^+$ implanted PDs, and were repeated for an un-implanted diode (having identical *p* and *n* implants and contact metallization as the $Zn^+$ implanted PD) of the same length. Using these measurements, the effects of parasitic absorption in the Al contacts is taken into account by calculating the modal absorption coefficient, $α_{eff} = -T/L$ [dB/cm], where *T* is the power transmission in dB and $L = 250$ μm (Fig. 1**c**). By subtracting the parasitic absorption coefficient $α_{eff,par}$, measured from an un-implanted diode, from the absorption coefficient of a $Zn^+$ implanted PD, $α_{eff,tot}$, the absorption coefficient due to defects shown in Fig. 3 is determined: $α_{eff,Zn} = α_{eff,tot} - α_{eff,par}$. The contact absorption becomes dominant for $λ > 2.325$ μm (see SI), resulting in the loss of responsivity with increasing wavelength seen in Fig. 2**d**.

**Acknowledgements:**

The authors thank Michael W. Geis and Steven J. Spector at MIT Lincoln Laboratory for waveguides and fabrication advice, and Sander Mann for a careful reading of the manuscript. The authors acknowledge support from the Columbia Optics and Quantum Electronics IGERT under NSF grant DGE-1069420. Research carried out in part at the Center for Functional Nanomaterials, Brookhaven National Laboratory, which is supported by the U.S. Department of Energy, Office of Basic Energy Sciences, under Contract No. DE-AC02-98CH10886.

**Author contributions:**

R. R. G. and B. S. performed the photocurrent and transmission measurements with guidance and supervision from W. M. J. G. Additional transmission measurements were performed by R. R. G and N. O. with guidance and supervision by K. B. and R. M. O. $Zn^+$ implantation was performed by B. S. and H. B. All authors contributed to data analysis and writing of the manuscript.

## Supplementary Information

**Measurement setup:**

Our measurement setup is shown in Fig. S1. An external cavity $Cr^{2+}$:ZnSe tunable laser (IPG Photonics) passes through an isolator (ISO) and is free-space coupled to a single-mode fiber (SMF) using a fiber collimator (FC). The SMF is also connected to a polarization rotator (PR) followed by a single-mode lensed-tapered fiber (LTF) designed for a 2.5 μm spot-size and working distance of 14 μm at $\lambda = 1.55$ μm (Oz optics). The LTF couples into a 5 μm × 220 nm fan-out tapered silicon waveguide, which adiabatically decreases in width over a length of 450 μm to a 520 nm × 220 nm channel waveguide. The PR is used to ensure that only the quasi-TE (QTE) mode of the waveguide is being excited, which produces maximum photocurrent as the overlap with the $Zn^+$ implanted region is larger than that of the quasi-TM mode. The channel waveguide adiabatically tapers into a rib/ridge waveguide with a 50 nm ridge and the same channel dimensions over a length of 100 μm, which then enters the $Zn^+$ implanted-silicon-waveguide photodiode. Reverse-bias voltage is applied and current is measured using a Keithley 2400 Source/Meter through electrical probes, which are landed on Al contact pads. Power transmitted through the photodiode is out-coupled in an identical manner to the input coupling, and passes through an SMF patch cable to a Yokogawa AQ6375 optical spectrum analyzer.

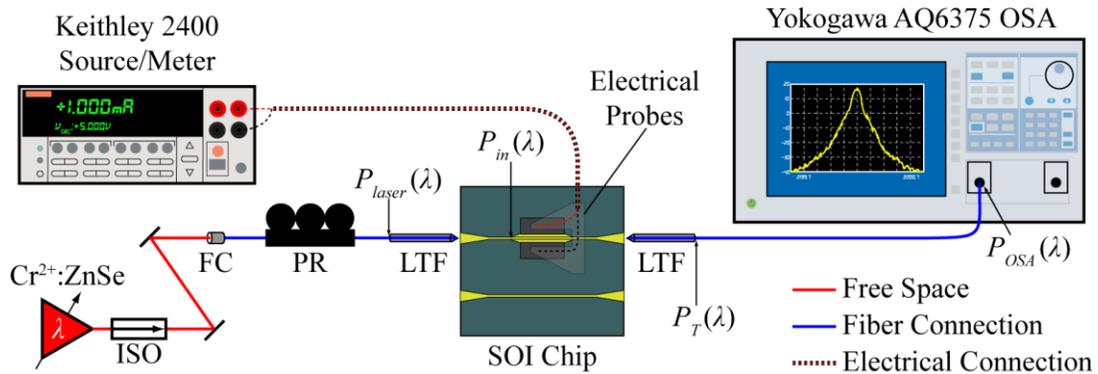

**Fig. S1** | Experimental setup. ISO = isolator, FC = fiber collimator, PR = polarization rotator, LTF = lensed tapered fiber.

**Determination of error bars:**

The total fiber-to-fiber transmission loss is measured by aligning two lensed-tapered fibers at their focal distance, and is defined in dB as $P_{F2F} = P_{laser} - P_{OSA}$, where $P_{laser}$ and $P_{OSA}$ are defined in Fig. S1. This loss is averaged over four measurements to determine the mean and standard deviation, $\bar{P}_{F2F}$ and $\sigma_{F2F}$, respectively. The transmission loss through the connecting fiber in dB is found as $P_{CF} = P_T - P_{OSA}$, and is also averaged over four measurements to find the mean and standard deviation $\bar{P}_{CF}$, $\sigma_{CF}$. The loss due to transmission through a single LTF is calculated as $\bar{P}_{LTF} = \frac{1}{2}(\bar{P}_{F2F} - \bar{P}_{CF})$ and the standard deviation is found by propagation of error $\sigma_{LTF} = \frac{1}{2}\sqrt{\sigma_{F2F}^2 + \sigma_{CF}^2}$.

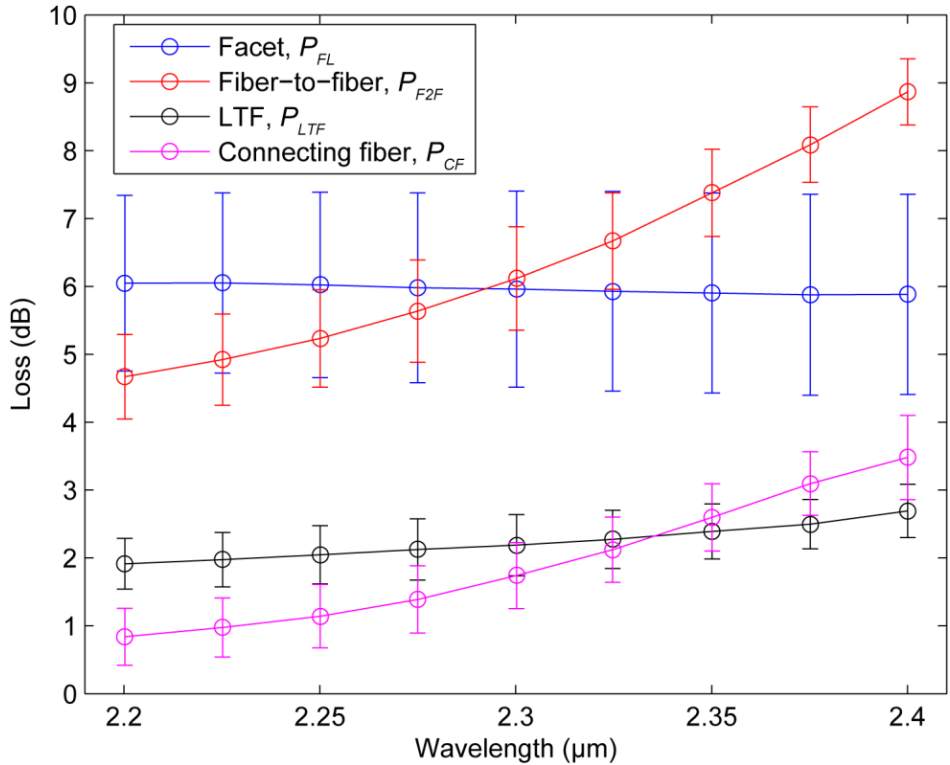

**Fig. S2 |** Measured system losses with error bars.

Facet loss from coupling between the LTF and a 5 μm × 220 nm fan-out taper was determined by averaging over 14 measurements of $\bar{P}_{FL} = \frac{1}{2}(\bar{P}_{laser} - \bar{P}_{OSA} - \bar{P}_{F2F})$, resulting in the mean, $\bar{P}_{FL}$, and standard deviation $\sigma_{FL} = \frac{1}{2}\sqrt{\sigma_{F2F}^2 + \text{var}(P_{laser} - P_{OSA})}$ shown in Fig. S2.

To determine error bars for the responsivity, the on-chip power is determined in milliwatts, $\bar{P}_{laser}[mW] = 10^{(P_{laser} - \bar{P}_{FL} - \bar{P}_{LTF})/10}$ and the propagation of error is approximated by a Taylor series expansion $\sigma_{laser}[mW] \approx \bar{P}_{laser}\frac{\ln(10)}{10}\sqrt{\sigma_T^2 + \sigma_{FL}^2 + \sigma_{LTF}^2}$. The mean responsivity is calculated as $\bar{R} = \frac{I_{ph}}{\bar{P}_{in}}$ resulting in a standard deviation of $\sigma_R \approx \bar{R}\frac{\sigma_{in}}{\bar{P}_{in}}$.

The error bars for the modal absorption coefficient in Fig. 3 of the main text are determined by measuring the optical power transmission $\bar{T} = P_{in} - 2\bar{P}_{FL} - \bar{P}_{F2F} - P_{OSA}$ through a 250 μm long PD, with the modal absorption coefficient being defined as $\bar{\alpha} = \frac{-\bar{T}}{L}$. The standard deviation is determined by propagation of uncertainty to be $\sigma_\alpha = \left(\sqrt{(2\sigma_{FL})^2 + \sigma_{F2F}^2}\right)/L$. This calculation is repeated with measurements from an un-implanted diode, and the modal absorption coefficient due to Zn⁺ implantation is defined as $\bar{\alpha}_{eff,Zn} = \bar{\alpha}_{eff,tot} - \bar{\alpha}_{eff,par}$ with standard deviation $\sigma_{\alpha,Zn} = \sqrt{\sigma_{\alpha,tot}^2 + \sigma_{\alpha,par}^2}$.

**Estimation of Zn and defect concentrations after implantation:**

The SRIM calculated dose profile is shown in Fig. S3. The averaged Zn concentration in the waveguide is estimated by integrating the Gaussian fit over the 220 nm length of the silicon device layer, and multiplying by the dose. The ion stopping range is adjusted by changing the acceleration voltage. For the PDs explored here, an acceleration voltage of 260 keV was used to place the stopping range in the center of the silicon device layer such that the fundamental quasi-TE mode of the SiWG (Fig 1**b**, main text) had maximum overlap with the created trap-states.

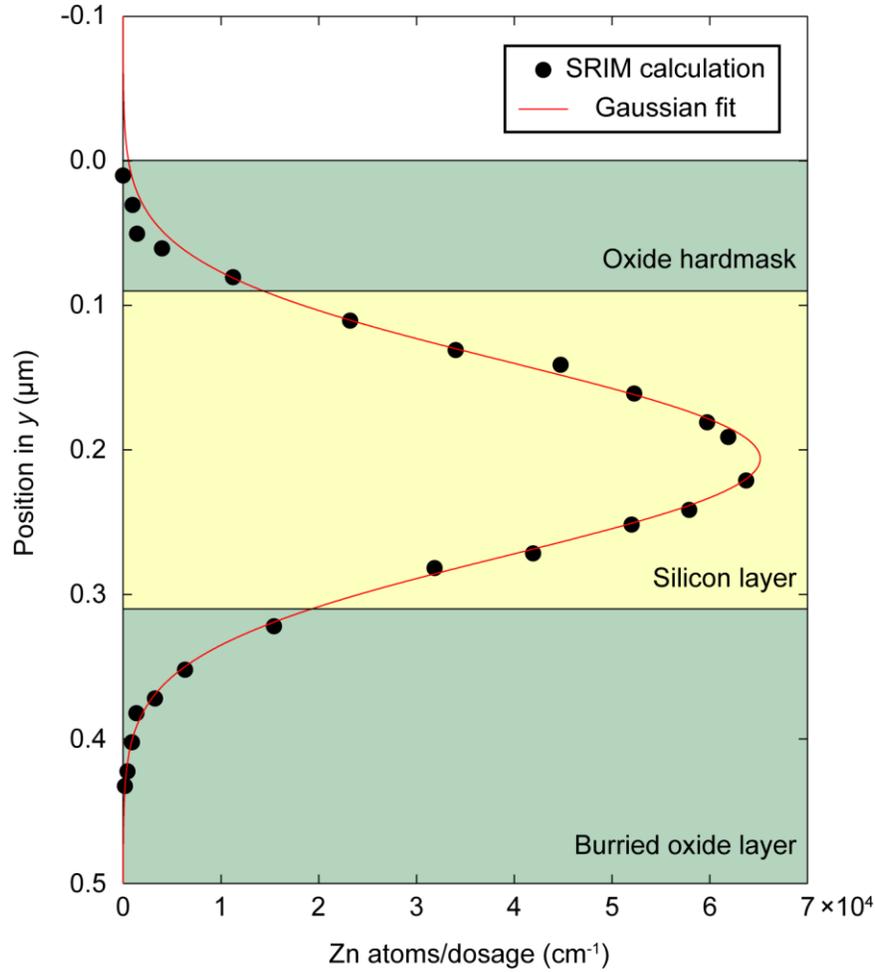

**Fig. S3** | SRIM calculated $Zn^+$ implantation profile for an acceleration voltage of 260 keV.

The amorphization threshold is taken to be the dose for which the concentration of implantation-induced lattice defects accounts for 10% of the total available Si lattice sites. This corresponds to a defect concentration of $5 \times 10^{21}$ cm$^{-3}$ for Si. Based on the SRIM calculations the threshold dose for amorphization is $2 \times 10^{13}$ cm$^{-2}$; however, these calculations do not take thermal effects into account. Thus, the SRIM estimation is an underestimate of the amorphization threshold dose. Even so, this conservative estimate shows that the doses explored here are below the amorphization threshold.

**Parasitic contact absorption:**

The transmission measurements used to find the modal absorption coefficients in Fig. 3 of the main text can also be used to determine the fraction of power absorbed in the PD that is due to Zn$^+$ implantation. Using the parasitic absorption coefficient $\alpha_{eff,par}$, measured from an un-implanted diode, and the absorption coefficient of a $p$-Si:Zn-$n$ PD, $\alpha_{eff,tot}$, the fraction of power absorbed by Zn$^+$ implantation can be found by noting that $P_{abs,Zn} = \alpha_{eff,Zn} P_{in} \int_0^L \exp(-\alpha_{eff,tot} z) \, dz = \frac{\alpha_{eff,Zn}}{\alpha_{eff,tot}} P_{in}[1 - \exp(-\alpha_{eff,tot} L)] = \frac{\alpha_{eff,Zn}}{\alpha_{eff,tot}} P_{abs,tot}$. The fraction of absorption due to Zn$^+$ implantation is shown in Fig. S4, and is determined as:

$$\frac{P_{abs,Zn}}{P_{abs,tot}} = \frac{\alpha_{eff,Zn}}{\alpha_{eff,tot}} = 1 - \frac{\alpha_{eff,par}}{\alpha_{eff,tot}}.$$

The error bars have been determined as described in the previous section.

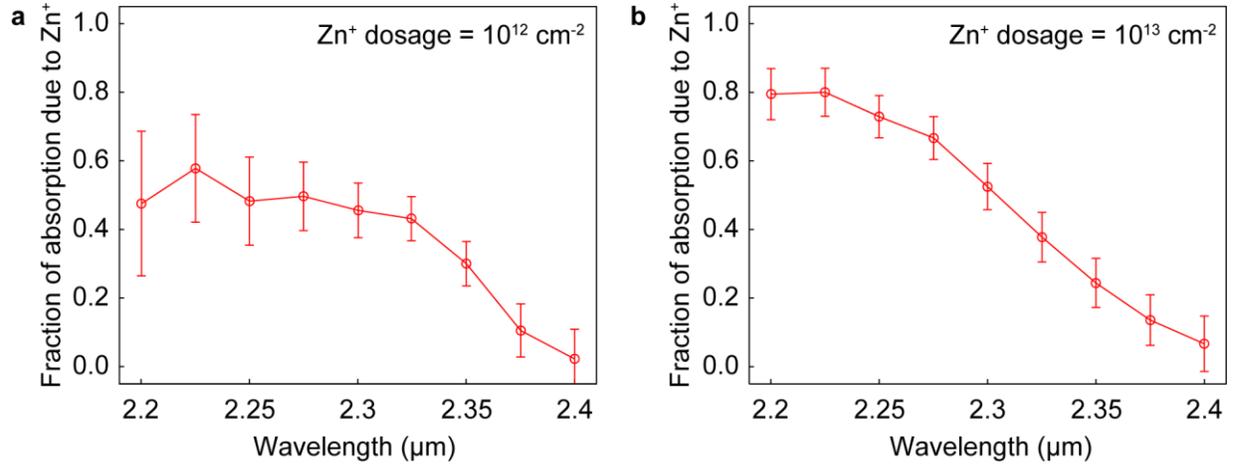

**Fig. S4** | Fraction of total power absorption due to Zn+ implantation, $\alpha_{eff,Zn}/\alpha_{eff,tot}$, for **a** Zn$^+$ dose = $10^{12}$ cm$^{-2}$ and **b** Zn$^+$ dose = $10^{13}$ cm$^{-2}$.

**Data from 3 mm length PDs**

Measurements of 3 mm-length Zn$^+$-implanted SiWG PDs are shown in Fig. S5 for implantation doses of $10^{12}$ cm$^{-2}$ and $10^{13}$ cm$^{-2}$. Photocurrent was measured beyond a wavelength of 2.35 μm; however, transmitted power could not be measured at these wavelengths due to the longer PD length, and thus error bars could not be determined for the responsivity.

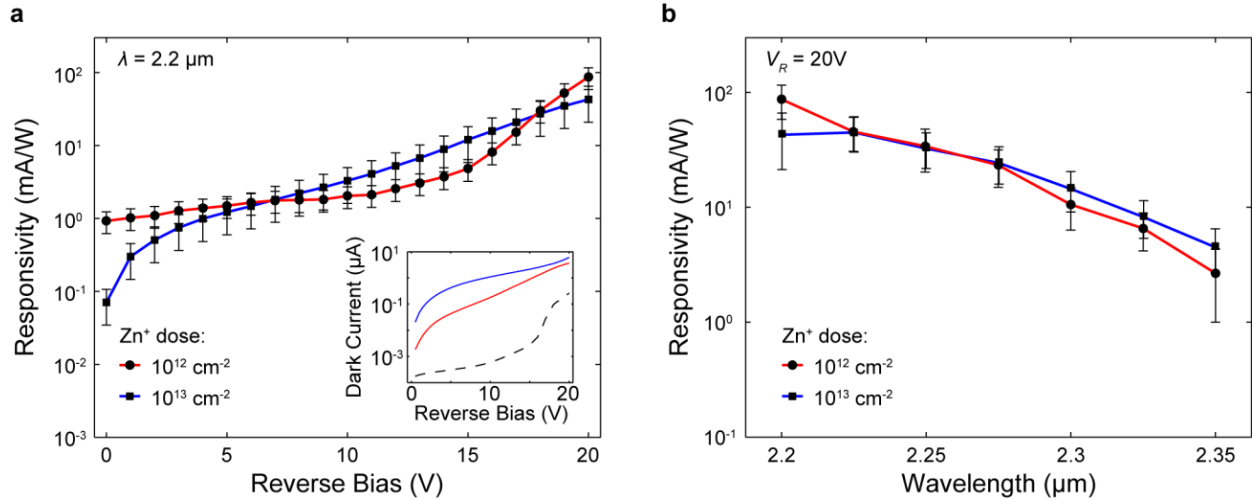

**Fig. S5 | a** Responsivity as a function of reverse bias voltage for 3 mm length PDs. Inset: dark current as a function of reverse bias voltage for an un-implanted diode (black dashed line) and PDs with Zn$^+$ doses of $10^{12}$ cm$^{-2}$ (red solid line) and $10^{13}$ cm$^{-2}$ (blue solid line). **d** Responsivity as a function of wavelength for 3 mm length PDs. Photocurrent was measurable above 2.35 μm; however, the transmitted power was not sufficient for determining error bars for the responsivity.